\documentclass[conference]{IEEEtran}
\IEEEoverridecommandlockouts

\usepackage{cite}
\usepackage{amsmath,amssymb,amsfonts}
\usepackage{algorithmic}
\usepackage{graphicx}
\usepackage{textcomp}
\usepackage{booktabs}
\usepackage{url}
\usepackage{xcolor}
\usepackage{chngcntr}
\def\BibTeX{{\rm B\kern-.05em{\sc i\kern-.025em b}\kern-.08em
    T\kern-.1667em\lower.7ex\hbox{E}\kern-.125emX}}
\begin{document}
\IEEEoverridecommandlockouts

\title{VHDL-Eval: A Framework for Evaluating Large Language Models in VHDL Code Generation
}



\author{
\IEEEauthorblockN{Prashanth Vijayaraghavan}
\IEEEauthorblockA{
\textit{IBM Research}\\
San Jose, CA 95120\\
prashanthv@ibm.com}
\and
\IEEEauthorblockN{Luyao Shi}
\IEEEauthorblockA{
\textit{IBM Research}\\
San Jose, CA 95120\\
luyao.shi@ibm.com}
\and
\IEEEauthorblockN{Stefano Ambrogio}
\IEEEauthorblockA{
\textit{IBM Research}\\
San Jose, CA 95120\\
stefano.ambrogio@ibm.com}
\and
\IEEEauthorblockN{Charles Mackin}
\IEEEauthorblockA{
\textit{IBM Research}\\
San Jose, CA 95120\\
charles.mackin@ibm.com
}
\and
\IEEEauthorblockN{Apoorva Nitsure}
\IEEEauthorblockA{
\textit{IBM Research}\\
San Jose, CA 95120\\
Apoorva.Nitsure@ibm.com}
\and
\IEEEauthorblockN{David Beymer}
\IEEEauthorblockA{
\textit{IBM Research}\\
San Jose, CA 95120\\
beymer@us.ibm.com
}
\and 
\IEEEauthorblockN{Ehsan Degan}
\IEEEauthorblockA{
\textit{IBM Research}\\
San Jose, CA 95120\\
edehgha@us.ibm.com
}
}

\maketitle
\IEEEpeerreviewmaketitle

\begin{abstract}
With the unprecedented advancements in Large Language Models (LLMs), their application domains have expanded to include code generation tasks across various programming languages. While significant progress has been made in enhancing LLMs for popular programming languages, there exists a notable gap in comprehensive evaluation frameworks tailored for Hardware Description Languages (HDLs), particularly VHDL. This paper addresses this gap by introducing a comprehensive evaluation framework designed specifically for assessing LLM performance in VHDL code generation task. We construct a dataset for evaluating LLMs on VHDL code generation task. This dataset is constructed by translating a collection of Verilog evaluation problems to VHDL and aggregating publicly available VHDL problems, resulting in a total of 202 problems. To assess the functional correctness of the generated VHDL code, we utilize a curated set of self-verifying testbenches specifically designed for those aggregated VHDL problem set. We conduct an initial evaluation of different LLMs and their variants, including zero-shot code generation, in-context learning (ICL), and Parameter-efficient fine-tuning (PEFT) methods. Our findings underscore the considerable challenges faced by existing LLMs in VHDL code generation, revealing significant scope for improvement. This study emphasizes the necessity of supervised fine-tuning code generation models specifically for VHDL, offering potential benefits to VHDL designers seeking efficient code generation solutions. 

\end{abstract}


\section{Introduction}

The rapid evolution of Large Language Models (LLMs) has propelled their widespread adoption across various Natural Language Processing (NLP) tasks, expanding their applicability into diverse domains such as Electronic Design Automation (EDA) and hardware design \cite{huang2021machine,rapp2021mlcad,thakur2023autochip,zhong2023llm4eda,liu2023rtlcoder,delorenzo2024make,chang2023chipgpt,blocklove2023chip,liu2023chipnemo}. Researchers have investigated how LLMs can augment EDA tools, enabling tasks like generating scripts to control EDA tools \cite{wu2024chateda,liu2023chipnemo}, integrating conversational LLMs in hardware design \cite{chang2023chipgpt,blocklove2023chip}, and addressing hardware security concerns \cite{nair2023generating,kande2023llm}. Notable studies, such as Chip-Chat \cite{blocklove2023chip} and Chip-GPT \cite{chang2023chipgpt}, have discussed challenges and opportunities in hardware design based on LLMs, highlighting the superior performance of certain models like ChatGPT over open-source alternatives. Particularly intriguing is the potential for LLMs to automatically generate RTL design based on natural language instructions, promising significant improvements in streamlining the hardware design process \cite{chang2023chipgpt,blocklove2023chip,liu2023chipnemo,lu2024rtllm,liu2023rtlcoder,thakur2023autochip}.

\begin{figure}[!ht]
    \centering
    \includegraphics[width=0.45\textwidth]{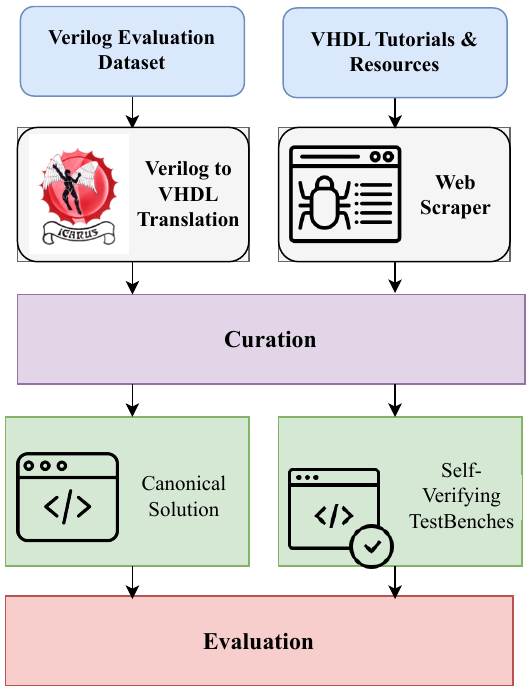}
    \caption{Overview of our Evaluation Framework for VHDL Code Generation}
    \label{fig:problem_st}
\end{figure}

Recently, there has been a surge in interest in generating Verilog code using fine-tuned LLMs, exemplified by Thakur et al.'s exploration using a fine-tuned CodeGen model \cite{thakur2023benchmarking,thakur2023verigen} for specific design tasks. However, most existing works in hardware design have focused on generating design RTL in Verilog, leaving a noticeable void in the realm of RTL design generation in VHDL (VHSIC Hardware Description Language). While models like Codegen \cite{nijkamp2022codegen,nijkamp2023codegen2} show promise in understanding VHDL code, the absence of a tailored evaluation framework optimized for VHDL poses a significant challenge in assessing LLM efficacy in this domain. VHDL's strict syntactical and semantic rules demand precise evaluation methodologies. Therefore, our research aims to develop such a framework specifically to assess the syntactic and functional correctness of VHDL code generated by LLMs.

To achieve this goal, we curated a dataset focused on assessing LLMs in VHDL code generation tasks. This dataset was meticulously constructed by translating a corpus of Verilog evaluation problems from a prior work \cite{liu2023verilogeval} into VHDL and collecting publicly available VHDL problem sets from tutorials, ensuring a diverse and comprehensive testbed for evaluating LLM performance. With a corpus comprising 202 problems spanning a spectrum of complexity levels, our dataset aims to comprehensively assess the robustness and adaptability of LLMs in the VHDL landscape. Integral to our evaluation framework is the incorporation of self-verifying testbenches designed to validate the functional correctness of the generated VHDL code. Usually, Verilog code completions are tested for functional correctness by comparing simulation outputs with golden solutions. In contrast, our self-verifying testbenches evaluate VHDL code generated by LLMs against test cases that the ground truth solutions pass, providing a clear measure of performance. Figure \ref{fig:problem_st} illustrates an overview of the evaluation framework for VHDL code completion.

To systematically evaluate the generated code, we prioritize two primary goals: ensuring syntactic correctness and validating functional correctness through self-verifying testbenches. In our pursuit of evaluating LLMs in VHDL code generation, we explore a diverse array of methodologies, including zero-shot code generation, in-context learning (ICL) and Parameter-Efficient Fine-Tuning (PEFT). Through a comprehensive evaluation of different LLMs and their variants, our research uncovers the formidable challenges confronting existing LLMs in the domain of VHDL code generation and underscores the critical importance of tailoring code generation models specifically for VHDL. By elucidating the challenges and opportunities in VHDL code generation, this study highlights the need for refinement and optimization of LLMs, thereby offering tangible benefits to VHDL designers striving to streamline and expedite the hardware design process. Our key contributions are summarized as follows:
\begin{itemize}
\item Construction of a curated dataset comprising 202 problems sourced from Verilog evaluation problems transposed into VHDL, combined with publicly available VHDL problem sets.
\item Development of a benchmarking framework incorporating self-verifying testbenches to assess the functional correctness of generated VHDL code.
\item Exploration of diverse methodologies including zero-shot code generation, in-context learning (ICL), and Parameter-efficient fine-tuning (PEFT) approaches to evaluate different LLMs and their variants. Our findings reveal significant scope for improvement in LLM performance on VHDL code generation tasks and emphasize the necessity of tailoring code generation models specifically for VHDL.
\end{itemize}

\section{Dataset}

We introduce the VHDL-Eval\footnote{The dataset will be published soon after getting clearance.} benchmark, designed for evaluating VHDL code generation models. This dataset comprises 202 code problems tailored for VHDL. It is constructed by translating problems from the Verilog-Eval dataset \cite{liu2023verilogeval}, which originally consisted of evaluation problems for Verilog, into VHDL. Additionally, we scrape publicly available VHDL problem sets from tutorials to enrich the dataset. Table \ref{tab:stats} provides the dataset statistics of the aggregated dataset.


\begin{table}[]
\centering
\caption{Dataset Statistics of our aggregated VHDL-Eval dataset.\label{tab:stats}}
\begin{tabular}{@{}lr@{}}
\toprule
\multicolumn{2}{c}{Dataset Statistics}                             \\ \midrule
\multicolumn{1}{l|}{\#Verilog-Eval Translated to VHDL Code}     & 113 \\
\multicolumn{1}{l|}{\#VHDL Code from Tutorials}              & 131 \\
\multicolumn{1}{l|}{\#Total VHDL Code after Post-Processing} & 202 \\
\multicolumn{1}{l|}{\#Avg. lines in VHDL Canonical Solutions} & 33.5 \\
\multicolumn{1}{l|}{\#Avg. Test Cases in Testbenches}       & 4.82 \\ \bottomrule
\end{tabular}
\end{table}

\subsection{Verilog-Eval}

The Verilog-Eval dataset primarily consists of 156 problems sourced from HDLBits \cite{liu2023verilogeval}. However, it cannot be directly applied to systematically evaluate the performance of VHDL code generation models. To address this limitation, we propose the development of a VHDL variant of Verilog-Eval. For each problem in Verilog-Eval, we translate the corresponding canonical solution into VHDL code snippets using the ICARUS Verilog tool. We discard samples that cannot be translated by the tool. Furthermore, we manually craft problem descriptions and self-verifying testbenches, each containing a varying number of test cases for these problems. We show an example of a half-adder in Figure \ref{vhdl_eval} in the Appendix.

\subsection{VHDL Tutorials}
\label{sec:tutorial}
We also incorporate problem statements from publicly available VHDL tutorials to augment the dataset. By scraping various VHDL tutorial websites\footnote{\url{https://www.fpga4student.com/p/vhdl-project.html, http://esd.cs.ucr.edu/labs/tutorial/, http://www.pldworld.com/_hdl/2/-seas.upenn.edu/_ese201/vhdl/vhdl_primer.html}}, we gather problem statements with varying levels of complexity. We utilize the Scrapy web framework, a Python tool for webpage crawling, to extract relevant data from these tutorials. Adhering to the usage policies of these tutorial pages, we primarily extract problem statements for our dataset.

\begin{figure}[!ht]
    \centering
    \includegraphics[width=0.5\textwidth]{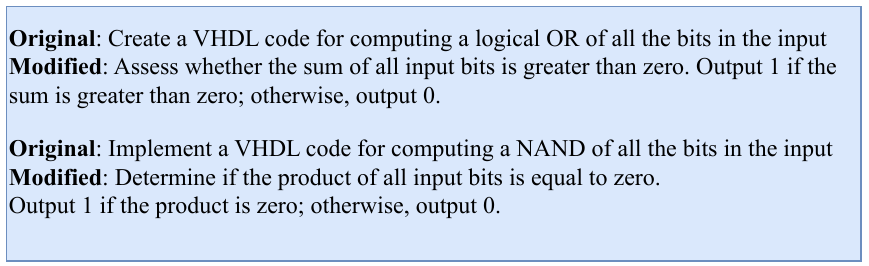}
    \caption{Example reframing of a standard problem as prompt.}
    \label{fig:problem_reframe}
\end{figure}

\paragraph{Post-Processing Module}
The aggregated datasets undergo a post-processing step. This module parses webpages to extract problem statements and filters out duplicate statements overlapping with the Verilog-Eval dataset. We, then, manually generate compilable VHDL canonical solutions and testbenches. These solutions and testbenches adhere to the template of the VUnit Testing framework (explained in Section \ref{sec:vunit}). To mitigate potential benchmark leakage issues \cite{zhou2023don}, where evaluation data is inadvertently used for model training, we reframe problem statements to prevent direct reuse. For instance, we modify basic logic gate implementations to involve n-bit implementations or alter the problem context where logic gates are part of a larger problem solution. An example is illustrated in Figure \ref{fig:problem_reframe}. This approach ensures a comprehensive dataset of VHDL code samples with varying complexity levels.

\subsection{Dataset Format}
Inspired by the HumanEval dataset, we format our dataset as follows:

\begin{itemize}
    \item \textit{task\_id}: Indicates the source followed by a problem-specific unique identifier. For example, verilog\_eval/always\_case refers to a problem from the Verilog-Eval dataset, while tutorial/ripple\_carry\_adder denotes a problem scraped from a tutorial.
    \item \textit{declaration}: Function/entity declaration including necessary libraries or packages.
    \item \textit{problem\_statement}: Brief description of the problem.
    \item \textit{description}: Detailed explanation specifying the functionality along with example input/output.
    \item \textit{prompt}: Commented problem statement concatenated with function/entity declaration.
    \item \textit{canonical\_solution}: Verified solution to the problem.
    \item \textit{testbench}: Test program including test cases.
\end{itemize}

\section{Experimental Setup}
\subsection{VHDL Testing Framework}
\label{sec:vunit}
VUnit\footnote{https://vunit.github.io/index.html} is a free and open source (FOSS) unit testing framework for VHDL that supports ModelSim, Rivera-PRO, GHDL, and Active-HDL. In our work, we primarily use VUnit with GHDL, which is a FOSS simulator. VUnit provides a check package for making tests self-checking and the test runner is responsible for the execution. The test runner will scan for source files and tests, figure out their dependencies, compile them and then run the selected tests as specified from command line. 
\subsection{Benchmark Models}
In the experiment, we evaluate different publicly available LLMs and their variants with our proposed VHDL code generation benchmark:
\begin{itemize}
    \item Codegen2.5, which is a 7B autoregressive language model for program synthesis. This model is trained on StarCoderData for 1.4T tokens.
    \item Codellama-34b-instruct \& Codellama-70b-instruct, which are  34B and 70B parameter instruction-tuned variants of Code Llama, built on top of Llama 2 and designed for general code synthesis and understanding.
    \item Granite-20b-instruct \footnote{\url{https://huggingface.co/ibm-granite/granite-20b-code-instruct}}, which is a 20B parameter model fine tuned from Granite-20B-Code-Base on a combination of permissively licensed instruction data to enhance instruction following capabilities including logical reasoning and problem-solving skills.
    \item In-Context Learning:  We leverage the in-context learning ability of LLMs by placing ``training'' data into a prompt preceding each test prompt. We experiment with ``Codellama-70b-instruct'' model under ICL setup with k=5 examples due to the limits on the context length.
    \item QLoRA: Recently, low-rank adaptation of specific model parameters has enabled extremely parameter-efficient finetuning of LMs. In our work, we finetune CodeGen2.5 using the LoRA implementation from HuggingFace PEFT \cite{mangrulkar2022peft} and the 8-bit AdamW from bitsandbytes \cite{dettmers2023case}. We train with rank $r = 16$ adapters, $\alpha = 32$, and dropout $p = 0.1$.  We aggregate 1000 samples containing a short code description and their corresponding VHDL code from Github VHDL data obtained using its API\footnote{\url{https://docs.github.com/en/rest?apiVersion=2022-11-28}}.
    
\end{itemize}

\subsection{Metrics}
To assess the functional correctness of a model, we utilize the \textit{Pass@k} metric, a widely employed measure in recent literature concerning code generation models \cite{chen2021evaluating}. This metric evaluates whether a model successfully solves a given problem by determining if any of the $k$ generated samples pass the unit tests associated with that problem. Formally, we use:
\begin{equation}
\text{{Pass@k}} := \mathbb{E}_{problems}\left[1 - \frac{{\binom{n-c}{k}}}{{\binom{n}{k}}}\right]
\end{equation}
where $n \geq k$ denotes the total number of samples generated per task, with $c$ being the count of samples that pass testing. It is essential to ensure that the number of generated samples $n$ is sufficiently large to yield low variance estimates for \textit{Pass@k}, as suggested in the literature.

\section{Results \& Discussion}
Table \ref{tab:eval} summarizes the result of our evaluation on different models using our VHDL-Eval Dataset.
\begin{table}[]
\centering
\caption{Evaluation Results of different code generating LLMs on our VHDL-Eval dataset. \label{tab:eval}}
\begin{tabular}{@{}llll@{}}
\toprule
\textbf{Models}            & \textbf{Pass@1} & \textbf{Pass@3} & \textbf{Pass@5} \\ \midrule
Codellama-70b-instruct     & 0.229           & 0.342           & 0.391           \\
Codellama-70b-instruct-ICL & \textbf{0.237}  & \textbf{0.349}  & \textbf{0.405}  \\
Codellama-34b-instruct     & 0.184            & 0.224           & 0.295           \\
CodeGen2.5                 & 0.072            & 0.110            & 0.190           \\
CodeGen2.5-QLoRa           & 0.095           & 0.129           & 0.215            \\
 Granite-20B-Code-Instruct & 0.085           & 0.146           & 0.219 \\
 Granite-20B-Code-Instruct-ICL & 0.106           & 0.195           & 0.248\\
\bottomrule
\end{tabular}
\end{table}

\subsection{Zero-shot Performance of Code-generating LLMs}

Table \ref{tab:eval} presents the zero-shot performance of three models in VHDL code generation: Codellama-70b-instruct, Codellama-34b-instruct, and Codegen2.5. Among these models, Codellama-70b-instruct stands out as the top performer, benefitting from its extensive parameter size which enables it to effectively capture patterns in VHDL code. Notably, Codellama-34b-instruct shows comparable Pass@1 performance to its 70b parameter counterpart, but the performance gap widens for Pass@3 and Pass@5. Additionally, the Granite-20b-instruct model's performance notably improves with ICL, particularly in the Pass@5 metrics. This suggests that while increasing the model size can lead to higher performance, there may still be opportunities for further improvement through exposure to more VHDL data and tailored fine-tuning. Despite its smaller size, Codegen2.5 demonstrates respectable performance in VHDL code generation, possibly due to its exposure to some VHDL data during training, highlighting its effective learning and generalization of VHDL code patterns.

\subsection{Effect of In-Context Learning (ICL)}

Our experiments with in-context learning (ICL) show marginal performance improvement for Codellama-70b-instruct and Granite-20b-instruct models. Even with a limited number of demonstrations ($k=5$), ICL enhances VHDL code generation. This suggests potential for further improvement by incorporating more meaningful code snippets into the context, but challenges remain due to context length limits and VHDL verbosity. Future research will explore these possibilities.

\subsection{Effect of Parameter-Efficient Finetuning}



Parameter-efficient finetuning of Codegen2.5 using QLoRa shows notable performance improvement with a limited parallel dataset ($\sim$1000 samples). This highlights the potential of low-rank adaptation techniques like QLoRa for VHDL code generation, especially when trained on specific VHDL data.

The varying performances of different models reveal a gap in VHDL code generation capabilities compared to other languages, emphasizing the importance of supervised fine-tuning with VHDL-specific data. This approach is key to bridging the performance gap and providing efficient VHDL code generation solutions. Future research should focus on this to enhance VHDL code generation and streamline the hardware design process.

\section{Limitations}

In this study, we curated a VHDL evaluation dataset by translating Verilog-Eval data to VHDL and integrating it with publicly available problem sets from tutorials. We manually created testbenches to evaluate the functional correctness of code generated by various models. These testbenches, while useful, may not cover all critical scenarios and might not be exhaustive or sufficient for all problem sets. Their adequacy varies with each problem's characteristics, and creating comprehensive testbenches for new tasks can be time-consuming and difficult to scale. The problem sets and their solutions are self-contained, lacking real-life design complexities. We used an open-source VUnit-based test framework for automatic evaluation, but there may be more scalable methods for assessing functional equivalence. The evaluations conducted in this study represent preliminary explorations of a few openly accessible models. Further experiments on more code generation models and exploring different hyperparameters could yield improved results, which we leave for future research. Lastly, despite efforts to prevent benchmark leakage as explained in Section \ref{sec:tutorial}, we cannot guarantee that the problem sets and code were not part of the original training set of the LLMs evaluated, given the scale of their training data.

\section{Conclusion}

In this paper, we aimed to bridge the gap in evaluation frameworks tailored for VHDL code generation using Large Language Models (LLMs). Introducing our curated evaluation dataset, VHDL-Eval, specifically designed for assessing LLM performance in VHDL code generation tasks, we compiled approximately 200 problems sourced from Verilog evaluation challenges translated into VHDL, along with publicly available VHDL problem sets. Our framework also incorporated self-verifying testbenches to ensure the functional correctness of the generated VHDL code. Through our evaluation, we explored various methodologies, including zero-shot code generation, in-context learning (ICL), and parameter-efficient fine-tuning (PEFT) approaches, to assess different LLMs and their variants. Our findings highlighted the significant challenges faced by existing LLMs in VHDL code generation, indicating substantial room for improvement. Despite the observed performance gap compared to languages with extensive training data, we noted marginal to respectable improvements in VHDL code generation tasks, even with limited parallel data, through the application of ICL and QLoRa approaches. This underscores the critical importance of supervised fine-tuning with more VHDL-specific data to bridge the performance gap and provide efficient VHDL code generation solutions for hardware design engineers. Thus, our study lays the foundation for future research endeavors aimed at refining and optimizing LLMs specifically for VHDL code generation. 

\bibliographystyle{IEEEtran}
\bibliography{conf}


\appendix
\setcounter{figure}{0}
\renewcommand\thefigure{\Alph{section}.\arabic{figure}}

\begin{figure*}[tb]
    \centering
    \includegraphics[width=\textwidth]{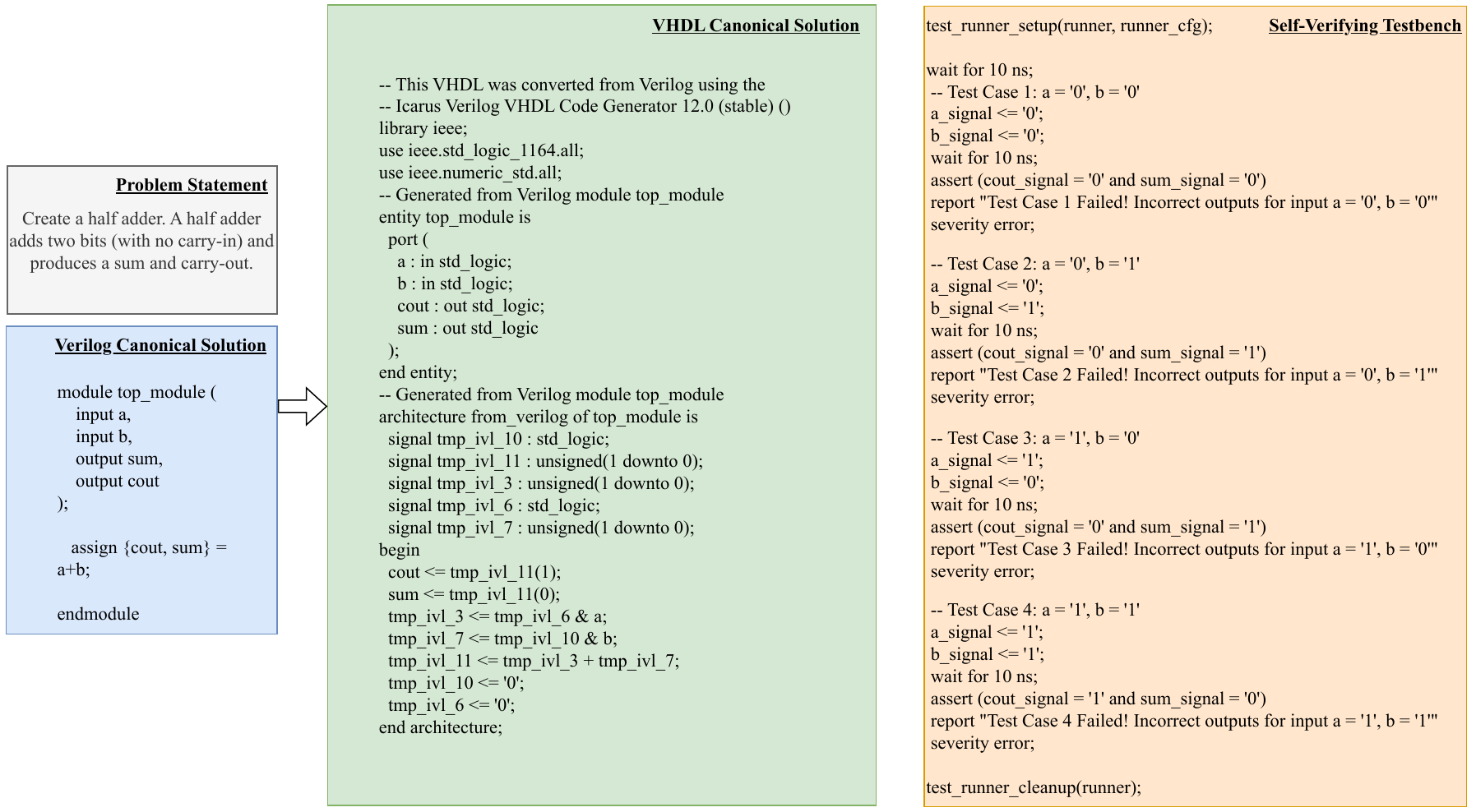}
    \caption{\textbf{Left}: Verilog half-adder problem statement and its canonical solution from the Verilog-Eval Dataset. \textbf{Center}: VHDL canonical solution for the half-adder obtained by translating the Verilog code using ICARUS Verilog tool. \textbf{Right}: Section of the self-verifying VHDL testbench for the half-adder, including test cases within the VUnit testing framework.}
    \label{vhdl_eval}
\end{figure*}

Figure \ref{vhdl_eval} shows an example of a half-adder from Verilog-Eval data set translated to VHDL, accompanied with 4 test cases.

\end{document}